\documentclass[preprint,12pt]{elsarticle}

\journal{Chaos, Solitons \& Fractals}

\usepackage[utf8]{inputenc}
\usepackage{graphicx}
\usepackage{multirow}
\usepackage{amsmath,amssymb,amsfonts}
\usepackage{mathrsfs}
\usepackage{xcolor}
\usepackage{textcomp}
\usepackage{booktabs}
\usepackage{microtype}

\biboptions{sort&compress}

\begin{document}

\begin{frontmatter}

\title{Dimensional reduction of higher-order interactions enables chaotic network reconstruction}

\author[aff1]{Jiahao He}
\ead{2410104045@mails.szu.edu.cn}

\author[aff1]{Xiaoqun Wu\corref{cor1}}
\ead{xqwu@szu.edu.cn}
\cortext[cor1]{Corresponding author}

\author[aff1,aff2]{Ziye Fan}
\ead{ziyefan@whu.edu.cn}

\author[aff3]{Jianwen Feng}
\ead{fengjw@szu.edu.cn}

\author[aff4]{Tingwen Huang}
\ead{huangtw2024@163.com}

\affiliation[aff1]{organization={Guangdong Provincial Key Laboratory of Visual Media and Multidimensional Intelligence, CSSE, Shenzhen University},
            city={Shenzhen},
            postcode={518060},
            country={China}}

\affiliation[aff2]{organization={School of Mathematics and Statistics, Wuhan University},
            city={Wuhan},
            postcode={430072},
            country={China}}

\affiliation[aff3]{organization={School of Mathematical Sciences, Shenzhen University},
            city={Shenzhen},
            state={Guangdong},
            postcode={518060},
            country={China}}

\affiliation[aff4]{organization={Faculty of Computer Science and Control Engineering, Shenzhen University of Advanced Technology},
            city={Shenzhen},
            postcode={518055},
            country={China}}

\begin{abstract}
Higher-order interactions create a combinatorial space of candidate dynamical terms, making network reconstruction difficult when the node dynamics, coupling functions and interaction topology are all unknown. Here we identify a regime in which weakly coupled higher-order chaotic networks admit a higher-order interaction reduction. Weak coupling largely preserves the statistical structure of intrinsic chaotic motion, while dense higher-order neighborhoods provide repeated sampling of the many weakly correlated multi-node inputs received by hub nodes. Consequently, the high-dimensional interaction field can be represented by a compact function of the hub state, together with a node-dependent offset and finite-neighborhood fluctuations. We exploit this reduction to learn dynamically relevant coupling forms and subsequently reconstruct directed interactions of different orders by sparse regression. Tests on higher-order chaotic networks identify an effective region of the examined coupling-parameter space in which interactions can be accurately recovered, covering the representative weak-coupling settings considered in this study. At larger network sizes and higher hub-neighborhood densities, the method maintains stable reconstruction performance and accurately recovers interaction topologies of different orders. Simulations on higher-order networks constructed from empirical social and cortical pairwise topologies further show that structural priors reduce data requirements and that the framework remains effective for heterogeneous chaotic dynamics. These results identify reduction of aggregate higher-order inputs as an interpretable route to dimensional compression and joint dynamical and structural reconstruction.
\end{abstract}


\end{frontmatter}

\section{Introduction}

Dynamical networks serve as a general framework for studying complex systems, with nodes representing individual units and edges capturing their interactions~\cite{boccaletti2006complex}. Structural and dynamical analyses of such networks underpin the study of system stability, collective behavior and evolutionary processes, while also informing systematic intervention and control~\cite{mao2025asymmetricLyapunov,mao2025finiteTimeTracking,wu2024multiplexSynchronization}. Within this broad class, weakly coupled chaotic networks---whose nodes exhibit chaotic local dynamics and interact weakly---represent an important but comparatively understudied regime. Such networks frequently arise in information propagation, ecological systems, and neural dynamics~\cite{newman2018networks}. Although weak coupling can support macroscopic collective behavior, it simultaneously introduces analytical challenges---such as instability of synchronous states and rapid decay of inter-node correlations---which limit the applicability of traditional similarity-based or strong-coupling-based approaches. It is therefore important to develop methods that reliably infer interactions in this weak-coupling regime. 

However, traditional network models typically assume that interactions occur exclusively between pairs of nodes. While this pairwise assumption captures many essential properties, it fails to characterize the multiscale interactions observed in real-world systems, especially in cases where higher-order interactions play a decisive role in shaping global dynamics~\cite{battiston2020networks,grilli2017higher}. Consequently, recent studies have shifted focus toward higher-order networks composed of chaotic oscillators~\cite{nijholt2022emergent} and the collective dynamics induced by higher-order interactions~\cite{skardal2020higher,battiston2021physics,li2024multidirectedCSF}.

Against this background, reconstructing the topology of higher-order weakly coupled networks holds significant theoretical and practical value. Over the past two decades, various reconstruction approaches have been proposed, with the majority focusing on pairwise network structures. Classical strategies include adaptive synchronization methods~\cite{liu2009structure,WOS:000526775400047,fan2022identifying}, of which only Ref.~\cite{fan2022identifying} specifically addresses higher-order interactions, and compressed sensing approaches~\cite{han2015robust,wang2020recovering,WOS:001018711800027,hang2020symmetryCSF}, both of which typically require prior knowledge of the nodal dynamics or the coupling functions. In real-world applications, however, only time-series observations of node states are usually available. This limitation has spurred the development of data-driven approaches. Deep learning methods~\cite{kipf2018neural,WOS:001139144400053,han2024collective,WOS:001630224100004} and Granger-causality-based techniques~\cite{wu2012inferring,WOS:000526775400049} have been applied to this task, but they often struggle with scalability and the curse of dimensionality in higher-order or high-dimensional systems. Related statistical inference frameworks have been developed to reconstruct interactions from binary-state observations~\cite{ma2020data,WOS:000673624500003,wang2022simplicialBinaryContagion}, while Bayesian mixture-prior approaches have been proposed for unsigned and signed networks~\cite{WOS:001571499200001}.
Despite these advances, most existing methods either focus on dyadic interactions or address higher-order structures only under specialized dynamical or observational settings, leaving a need for a more general reconstruction framework for weakly coupled higher-order systems.

To address these challenges, data-driven reconstruction algorithms such as ARNI~\cite{casadiego2017model} and the recent hypergraph reconstruction approach in~\cite{delabays2025hypergraph} have emerged. Dynamics-based methods have also been extended from single-layer reconstruction to multilayer decomposition and, more recently, to multiplex networks with coexisting pairwise and higher-order interactions~\cite{zhang2023multilayerCSF,huang2026multiplexCSF}. Meanwhile, there has been growing interest in recovering interpretable equations directly from data to reveal the mechanisms behind observed dynamics~\cite{kim2021integration,zhang2024deep,wang2025advancing}. In this direction, these ideas have been further extended to networked systems, where learned equations can uncover latent mechanisms underlying the formation and evolution of complex phenomena~\cite{gao2024learning,hu2025learning,yu2025discover}. 
However, a fundamental challenge persists: when both the network structure and the governing equations are entirely unknown, existing methods often struggle due to the combinatorial growth of the candidate function library and the lack of prior structural knowledge. Previous results for pairwise weakly coupled systems suggest that the dynamics of highly connected nodes may admit a compact effective-coupling description~\cite{topal2023reconstructing,pereira2020heterogeneously}. Whether such statistical compression survives in genuinely higher-order networks, however, remains unclear. A \(D\)-th-order interaction depends on the joint states of multiple neighbors, producing combinatorially many candidate terms and residual multi-node correlations that cannot be reduced to pairwise contributions. The central question is therefore whether aggregate higher-order interactions possess a structure that makes reconstruction tractable.

Here we identify a higher-order interaction reduction that emerges under weak coupling and dense higher-order connectivity. Although individual higher-order terms depend on multiple neighbor states, their aggregate contribution at a hub is represented by a compact function determined primarily by the hub state, up to a node-dependent constant and finite-neighborhood fluctuations. This effective-coupling representation compresses the functional search space before topology inference. Building on this observation, we develop a three-step framework that learns the intrinsic dynamics, identifies dynamically relevant coupling functions and reconstructs directed interactions of multiple orders from discretely sampled continuous-time trajectories. Numerical experiments on second- and third-order networks support the proposed reduction and the resulting reconstruction framework.
The remainder of this paper is organized as follows. Section~2 formulates the higher-order network reconstruction problem, derives the interaction reduction and presents the reconstruction method. Section~3 reports the numerical results for randomly generated higher-order networks and networks constructed using empirical pairwise topologies. Section~4 discusses the operating regime, limitations and possible extensions, and concludes the study.

\section{Methods}
\subsection{Higher-order network model and reconstruction problem}

We consider a higher-order dynamical network consisting of $N$ nodes.
The state of node $i$ is denoted by $x_i\in\mathbb{R}^m$.
For clarity, we present the derivation for the scalar case $m=1$; the vector-valued case follows analogously.
The dynamics of node $i$ are governed by its intrinsic dynamics and interactions of possibly different orders:
\begin{equation}
\begin{aligned}
\dot{x}_{i} =& \ f(x_{i}) 
+ \sigma_{1} \sum_{j_1=1}^{N} a_{i j_1}^{(1)} g^{(1)}(x_{i}, x_{j_1}) \\
&+ \sigma_{2} \sum_{j_1=1}^{N} \sum_{j_2=1}^{N} 
a_{i j_1 j_2}^{(2)} g^{(2)}(x_{i}, x_{j_1}, x_{j_2})  
+ \cdots ,
\label{eq:network_model} 
\end{aligned}
\end{equation}
where $f(x_i)$ denotes the intrinsic dynamics of an isolated node, and $g^{(D)}$ denotes the interaction function of order $D=1,2,\ldots$.
The directed higher-order structure is encoded by the adjacency tensor
$A^{(D)}=[a^{(D)}_{ij_1\cdots j_D}]$, where $a^{(D)}_{ij_1\cdots j_D}=1$ if the group of nodes $(j_1,\ldots,j_D)$ jointly influences node $i$, and $a^{(D)}_{ij_1\cdots j_D}=0$ otherwise.

We focus on diffusion-like interactions across all orders.
The corresponding higher-order adjacency tensors satisfy the generalized Laplacian constraint
\begin{equation}
a_{\underbrace{\scriptstyle i\cdots i}_{D \ \text{orders}}}^{(D)}
= -\sum_{\substack{j_1=1 \\ j_1 \neq i}}^{N} 
   \sum_{\substack{j_2=1 \\ j_2 \neq i}}^{N} 
   \cdots 
   \sum_{\substack{j_D=1 \\ j_D \neq i}}^{N} 
   a_{ij_1j_2\cdots j_D}^{(D)} .
\label{eq:laplacian_D}
\end{equation}
The interaction functions are further assumed to admit the separable diffusion-like form
\begin{equation}
g^{(D)}(x_i,\{\cdot\})
=
h^{(D)}(\{\cdot\})-h^{(D)}(x_i,\ldots,x_i),
\label{eq:separable_def}
\end{equation}
where $\{\cdot\}$ denotes the collection of states from the $D$ interacting nodes.

Given this model, we consider a data-driven reconstruction problem in which only
multivariate time series of node states are available:
\begin{equation}
\mathcal{X}
=
\left\{
x_i(t_k)\in\mathbb{R}^m
\,\middle|\,
i=1,\ldots,N,\; k=1,\ldots,T
\right\},
\label{eq:measurements}
\end{equation}
where \(\{t_k\}_{k=1}^{T}\) are the sampling instants.
The goal is to reconstruct both the governing dynamics and the underlying higher-order interaction structure from \(\mathcal{X}\).
Specifically, we aim to identify the intrinsic node dynamics and the multi-node interaction terms that determine the evolution of the system.

\subsection{Emergent reduction of higher-order interactions in weakly coupled chaotic networks}

Higher-order interactions would appear to obstruct compact reduction because each interaction depends on the joint state of several neighbors and cannot, in general, be decomposed into pairwise terms. Previous pairwise results motivate asking whether an analogous effective-coupling description survives in genuinely higher-order networks~\cite{topal2023reconstructing}. We find that, within a weak-coupling, dense-neighborhood regime, the aggregate higher-order contribution admits such a representation. Weak coupling keeps individual trajectories close to the statistical behavior of their intrinsic chaotic dynamics, whereas hub nodes collect many weakly correlated multi-node inputs. Their cumulative effect can therefore be captured by a compact effective coupling at the aggregate level.

This higher-order interaction reduction relies on two conditions.
First, the coupling strengths must be sufficiently weak that, after transients, each node largely preserves the statistical properties of its intrinsic chaotic motion.
In this regime, the isolated dynamics remain dominant relative to coupling-induced perturbations.
Second, the network must contain informative hub nodes with sufficiently dense pairwise and higher-order neighborhoods.
Such hubs provide repeated sampling of neighbor states, allowing the cumulative interaction to be represented by physical-measure averages up to finite-neighborhood fluctuations and residual correlations.

We denote the total interaction term acting on node \(i\), as defined by the higher-order network model, by
\begin{equation}
\begin{aligned}
H(x_i) =& \; \sigma_{1} \sum_{j_1=1}^{N} a_{ij_1}^{(1)} g^{(1)}(x_i, x_{j_1}) \\
&+ \sigma_{2} \sum_{j_1=1}^{N} \sum_{j_2=1}^{N} 
a_{ij_1j_2}^{(2)} g^{(2)}(x_i, x_{j_1}, x_{j_2})  
+ \cdots .
\label{eq:coupling_general}
\end{aligned}
\end{equation}
Under the assumptions above, the neighbor-dependent components of \(H(x_i)\) can be approximated by statistical averages, leading to an effective-coupling representation.

Substituting Eq.~\eqref{eq:separable_def} into Eq.~\eqref{eq:coupling_general} gives
\begin{equation}
\begin{aligned}
H(x_i)= {}&\; \sigma_{1} \sum_{j_1=1}^{N} a_{ij_1}^{(1)}
\left( h^{(1)}(x_{j_1})-h^{(1)}(x_i) \right) \\
&+ \sigma_{2} \sum_{j_1=1}^{N} \sum_{j_2=1}^{N} a_{ij_1j_2}^{(2)}
\left( h^{(2)}(x_{j_1},x_{j_2})-h^{(2)}(x_i,x_i) \right) \\
&+ \cdots .
\label{eq:coupling_separable}
\end{aligned}
\end{equation}
Using the generalized Laplacian constraint in Eq.~\eqref{eq:laplacian_D}, the self-dependent terms can be separated from the neighbor-dependent contributions:
\begin{equation}
\begin{aligned}
H(x_i)=&
\;\sigma_{1} a_{ii}^{(1)} h^{(1)}(x_i) 
+ \sigma_{2} a_{iii}^{(2)} h^{(2)}(x_i,x_i) \\
&+ \sigma_{1}\sum_{\substack{j_1=1\\ j_1\neq i}}^{N}
a_{ij_1}^{(1)} h^{(1)}(x_{j_1}) \\
&+ \sigma_{2} \sum_{\substack{j_1=1\\ j_1\neq i}}^{N}
 \sum_{\substack{j_2=1\\j_2\neq i}}^{N} 
 a_{ij_1j_2}^{(2)} h^{(2)}(x_{j_1},x_{j_2})
 + \cdots .
\label{eq:coupling_isolated}
\end{aligned}
\end{equation}

In the weak-coupling regime, the intrinsic chaotic motion remains dominant after transients.
We therefore describe the uncoupled system \(\dot{x}=f(x)\) by a physical invariant measure \(\mu\) supported on the chaotic attractor.
For any continuous observable \(\varphi\), this measure satisfies
\begin{equation}
\label{eq:physical_measure_def}
\lim_{T\to\infty}\frac{1}{T}\int_{0}^{T}\varphi(x(t;x_0))\,dt
=
\int \varphi(x)\,d\mu(x),
\end{equation}
for a set of initial conditions \(x_0\) in the basin of the attractor.
Thus, in the weakly coupled network, the instantaneous node states can be approximated by the reference statistics of the uncoupled chaotic system, up to small coupling-induced perturbations.

For a hub node with a sufficiently dense neighborhood, the neighbor-dependent sums in Eq.~\eqref{eq:coupling_isolated} aggregate a large number of weakly correlated contributions.
Their empirical averages can therefore be approximated by the corresponding \(\mu\)-averages.
For higher-order terms, this approximation further assumes that the joint empirical distribution of neighbor states can be approximated by products of the physical measure, up to finite-degree fluctuations and residual correlations.
Averaging the neighbor-state terms with respect to the physical measure gives
\begin{equation}
\begin{aligned}
H(x_i) \approx& \; 
\sigma_{1} a_{ii}^{(1)} h^{(1)}(x_i) 
+ \sigma_{2} a_{iii}^{(2)} h^{(2)}(x_i,x_i)
+\cdots  \\
&-\sigma_{1} a_{ii}^{(1)} 
\int h^{(1)}(x_{j_1})\, d\mu(x_{j_1}) \\
&-\sigma_{2} a_{iii}^{(2)} 
\iint h^{(2)}(x_{j_1},x_{j_2})\,
d\mu(x_{j_1})\,d\mu(x_{j_2}) 
-\cdots .
\label{eq:coupling_meanfield}
\end{aligned}
\end{equation}

The \(\mu\)-averaged terms in Eq.~\eqref{eq:coupling_meanfield} are approximately time-independent after transients.
Accordingly, the neighbor-induced contribution can be decomposed as \(C_i+\varepsilon_i(t)\), where \(C_i\) denotes the node-dependent constant offset induced by the averaged neighbor contributions and \(\varepsilon_i(t)\) captures finite-degree effects and residual correlations.
Substituting this decomposition into Eq.~\eqref{eq:coupling_meanfield} yields the reduced interaction form
\begin{equation}
\label{eq:coupling_final}
\begin{aligned}
H(x_i) \approx& \; 
\sigma_{1} a_{ii}^{(1)} h^{(1)}(x_i)
+ \sigma_{2} a_{iii}^{(2)} h^{(2)}(x_i, x_i)
+ \cdots \\
&+ C_i + \varepsilon_i(t).
\end{aligned}
\end{equation}
Equation~\eqref{eq:coupling_final} therefore gives the effective-coupling representation of the higher-order interaction reduction, in which \(C_i\) captures the mean neighbor contribution and \(\varepsilon_i(t)\) accounts for finite-neighborhood fluctuations. It is a statistical approximation to the aggregate interaction under the conditions stated above.

\subsection{Reconstruction framework based on higher-order interaction reduction}

The higher-order interaction reduction in Eq.~\eqref{eq:coupling_final} represents the aggregate coupling effect using a small number of effective functions, thereby avoiding a direct search over the full higher-order candidate space during topology inference. Building on this property, we develop a three-step reconstruction method that sequentially identifies the intrinsic node dynamics, effective coupling forms and multi-order interaction topology.

\subsubsection{Self-dynamics identification and node classification}

For each node \(i\), the observed time series is denoted by
\begin{equation}
X_i=\bigl[x_i(t_1),x_i(t_2),\ldots,x_i(t_T)\bigr].
\label{eq:observed_time_series}
\end{equation}
The time derivative \(\dot X_i\) is estimated from the sampled trajectory. We use sparse identification of nonlinear dynamics~\cite{brunton2016sindy} to represent the node self-dynamics as
\begin{equation}
\label{eq:self_sindy}
\dot{\mathbf X}_i=\Theta(X_i)\Xi_i ,
\end{equation}
where \(\Theta(X_i)\) is a candidate library containing constant, polynomial, trigonometric or other nonlinear functions as appropriate, and \(\Xi_i=[\xi_i^1,\ldots,\xi_i^p]^\top\) is the sparse coefficient vector. We estimate \(\Xi_i\) by sequential thresholded least squares: an initial least-squares solution is thresholded at \(\lambda\), the retained coefficients are refitted on the active columns of \(\Theta(X_i)\), and the thresholding--refitting cycle is repeated until the active set stabilizes. The nonzero entries of \(\Xi_i\) thereby identify the active terms in the autonomous dynamics of node \(i\).

The coefficient vector \(\Xi_i\) also provides a compact representation for comparing node dynamics.
For two nodes \(i\) and \(j\), we define the coefficient-space distance
\begin{equation}
\label{eq:coef_distance}
d_{ij}
=
\left(
\sum_{k=1}^{p}
\frac{1}{V_k}
|\xi_i^k-\xi_j^k|^2
\right)^{1/2},
\end{equation}
where \(V_k\) normalizes the scale of coefficient \(k\) across nodes.
Nodes that are close to many other nodes in coefficient space are regarded as peripheral nodes and are used to refine the intrinsic dynamics \(\hat f\).
Nodes that remain distant from most others are regarded as hub nodes, whose fitted dynamics contain stronger aggregate coupling effects and are therefore used for effective-coupling identification.

\subsubsection{Effective-coupling identification and library reduction}

For a hub node \(i\), the coupling-induced residual is extracted by subtracting the estimated intrinsic dynamics from the observed derivative:
\begin{equation}
\label{eq:extract_H}
\hat H(x_i(t))
=
\dot{x}_i(t)-\hat f(x_i(t)).
\end{equation}
Under the weak-coupling and dense-neighborhood conditions used in the reduction, the aggregated influence of neighboring nodes is approximated by low-dimensional functions of the hub state, such as \(\hat h^{(1)}(x_i)\), \(\hat h^{(2)}(x_i,x_i)\), and \(\hat h^{(D)}(x_i,\ldots,x_i)\). These functions are selected by applying the same sparse thresholding--refitting procedure to the extracted residual over a generic library of candidate nonlinear functions.

These identified component functions are extended to candidate neighbor states to form a reduced coupling library:
\begin{equation}
\label{eq:Theta_h}
\boldsymbol{\Psi}(\mathbf X)
=
\big[
\{\hat h^{(1)}(x_{j_1})\}_{j_1},
\{\hat h^{(2)}(x_{j_1},x_{j_2})\}_{j_1,j_2},
\ldots,
\{\hat h^{(D)}(x_{j_1},\ldots,x_{j_D})\}_{j_1,\ldots,j_D}
\big].
\end{equation}
For each hub node, the extracted coupling residual is fitted as
\begin{equation}
\label{eq:coupling_function_regression}
\hat{\mathbf H}_i
=
\boldsymbol{\Psi}(\mathbf X)\boldsymbol{\Gamma}_i,
\qquad
\hat{\mathbf H}_i
=
[\hat H(x_i(t_1)),\ldots,\hat H(x_i(t_T))]^\top .
\end{equation}
The sparse vector \(\boldsymbol{\Gamma}_i=[\gamma_i^1,\ldots,\gamma_i^q]^\top\) selects the active coupling forms from the reduced library. A full higher-order library would require enumerating cross-node terms whose number grows combinatorially with network size and interaction order. By retaining only the coupling forms supported by the hub dynamics, the reduced library lowers the regression dimension while preserving the functional information required for topology reconstruction.

\subsubsection{Higher-order topology reconstruction}

After the effective coupling functions have been identified, the interaction term acting on node \(i\) can be represented as
\begin{equation}
\label{eq:H_sum_neighbors_explicit}
H(x_i)
=
\sum_{D=1}^{\bar D}
\sigma_D
\sum_{\substack{j_1=1 \\ j_1\neq i}}^{N}
\cdots
\sum_{\substack{j_D=1 \\ j_D\neq i}}^{N}
a_{ij_1\cdots j_D}^{(D)}
h^{(D)}(x_{j_1},\ldots,x_{j_D}) .
\end{equation}
Discretizing Eq.~\eqref{eq:H_sum_neighbors_explicit} over the sampled trajectory yields the node-wise inverse problem
\begin{equation}
\label{eq:YPhiA_explicit}
\hat{\mathbf H}_i
=
\boldsymbol{\Phi}_i\mathbf{A}_i .
\end{equation}
Here, \(\boldsymbol{\Phi}_i\) is obtained by evaluating the selected coupling functions on all admissible neighbor tuples, and \(\mathbf{A}_i\) collects the unknown off-diagonal adjacency coefficients \(a_{ij_1}^{(1)}, a_{ij_1j_2}^{(2)},\ldots,a_{ij_1\cdots j_D}^{(D)}\).
Following the optimization-based formulation for higher-order coupled systems~\cite{malizia2024reconstructing}, we solve Eq.~\eqref{eq:YPhiA_explicit} node by node using sparse thresholded regression. The nonzero entries of \(\mathbf A_i\) determine the directed pairwise and higher-order interactions pointing to node \(i\). Repeating the procedure for all target nodes recovers the multi-order interaction structure; the preceding library reduction makes this inverse problem tractable without enumerating a full set of generic nonlinear functions for every neighbor tuple.

\section{Numerical results}

Reconstruction performance is evaluated from two complementary perspectives. The effective-coupling fit is quantified by the coefficient of determination \(R^2\), whereas topology recovery is evaluated by
\begin{equation}
\begin{gathered}
\mathrm{Precision}=\frac{TP}{TP+FP},
\qquad
\mathrm{Recall}=\frac{TP}{TP+FN},\\
F_1=\frac{2\,\mathrm{Precision}\,\mathrm{Recall}}
{\mathrm{Precision}+\mathrm{Recall}}.
\end{gathered}
\label{eq:topology_metrics}
\end{equation}
Here, \(TP\), \(FP\), and \(FN\) are the numbers of true positives, false positives, and false negatives, respectively. Because the reconstructed higher-order networks are sparse, \(F_1\) is used as the primary topology metric. The effective-coupling fit is measured by
\begin{equation}
R^2
=
1-
\frac{
\sum_{t=1}^{T}
\left[
\widehat{H}(x(t))-H(x(t))
\right]^2
}{
\sum_{t=1}^{T}
\left[
H(x(t))-\overline{H}
\right]^2
},
\label{eq:R2_H}
\end{equation}
where \(H(x(t))\) is the extracted coupling signal, \(\widehat{H}(x(t))\) is the fitted effective coupling function, and \(\overline{H}\) is the temporal mean of \(H(x(t))\).

\subsection{Randomly generated higher-order networks}

We first validate the proposed reconstruction framework on a randomly generated higher-order Chen network, where the ground-truth pairwise and higher-order interactions are known.
The network consists of \(N=50\) nodes, each governed by Chen chaotic dynamics.
The first-order adjacency matrix \(A^{(1)}=(a_{ij_1}^{(1)})\) is generated using the Barabási--Albert model with attachment parameter \(m=4\) and seed size \(m_0=m+1\).
On the same node set, second-order interactions are represented by a sparse third-order adjacency tensor \(A^{(2)}=(a_{ij_1j_2}^{(2)})\).
To generate informative higher-order neighborhoods, source nodes are selected from the top quartile of the first-order degree ranking.
For each selected hub node \(i\), node pairs \((j_1,j_2)\) are uniformly sampled under the constraints \(j_1\neq i\), \(j_2\neq i\), and \(j_1\neq j_2\), with the number of sampled higher-order interactions scaled proportionally to the first-order degree of the hub node.

The network dynamics are given by
\begin{equation}
\label{eq:chen_second}
\begin{cases}
\begin{aligned}
\dot{x}_{i} &= \alpha(y_i - x_i) 
+\sigma_{1} \sum_{j_1=1}^N a_{ij_1}^{(1)}(x_{j_1} - x_i) \\
&\quad + \sigma_{2}  \sum_{j_1=1}^N \sum_{j_2=1}^N 
a_{ij_1j_2}^{(2)}(x_{j_1}^2 x_{j_2} - x_i^3), \\[2pt]
\dot{y}_i &= (\gamma-\alpha)x_i - x_i z_i + \gamma y_i, \\[2pt]
\dot{z}_i &= x_i y_i - \beta z_i,
\end{aligned}
\end{cases}
\end{equation}
with standard chaotic parameters \(\alpha=35\), \(\beta=3\), and \(\gamma=28\).

\begin{figure*}[htbp]
    \centering
    \begin{minipage}[t]{0.47\textwidth}
        \textbf{a}\par\vspace{1mm}
        \centering
        \includegraphics[width=\linewidth]{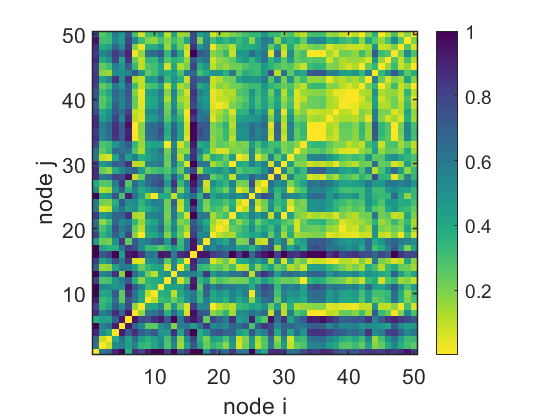}
    \end{minipage}
    \hspace{0.04\textwidth}
    \begin{minipage}[t]{0.47\textwidth}
        \textbf{b}\par\vspace{1mm}
        \centering
        \includegraphics[width=\linewidth]{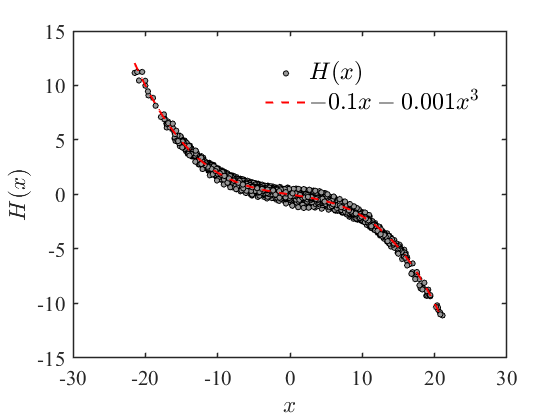}
    \end{minipage}

    \caption{
    Validation of the reconstruction procedure on a randomly generated higher-order Chen network.
    \textbf{a,} Normalized coefficient-space distance matrix between nodes, showing a clear separation between hub nodes and peripheral nodes.
    \textbf{b,} Reconstructed effective coupling function \(H(x_i)\) for a representative hub node and its polynomial fit.
    The clear node separation and accurate effective-coupling fit support the use of hub nodes for identifying reduced coupling functions.
    }
    \label{fig1}
\end{figure*}

In applying the reconstruction scheme, we first classify nodes using pairwise distances computed from their coefficient-space representations obtained by sparse regression.
For visualization, the distance matrix is normalized to the interval \([0,1]\).
As shown in Fig.~\ref{fig1}a, the resulting matrix clearly separates hub nodes from peripheral nodes.
Nodes with small pairwise distances share similar intrinsic dynamics and are used to estimate the self-dynamics, whereas hub nodes exhibit distinct coefficient-space behavior due to their stronger aggregate coupling inputs.

For a representative hub node \(i=1\), we set the coupling strengths as
\begin{equation}
\sigma_{1}=\frac{0.1}{|a_{11}^{(1)}|},
\qquad
\sigma_{2}=\frac{0.001}{|a_{111}^{(2)}|}.
\label{eq:normalized_coupling_strength}
\end{equation}
The extracted coupling term \(H(x_i)\) is then fitted using a candidate function set depending on the self-state \(x_i\).
As shown in Fig.~\ref{fig1}b, the effective coupling is well approximated by the shifted polynomial form
\begin{equation}
H(x_i) \simeq -0.1x_i - 0.001x_i^3 + C,
\label{eq:chen_effective_coupling}
\end{equation}
where \(C\) is approximately constant.
This result supports the proposed higher-order interaction reduction: although the original higher-order coupling depends on many neighbor states, the aggregate input received by an informative hub can be represented by a compact function of the hub state.

Building on this fitted coupling form, we construct a reduced candidate library containing the dominant interaction terms, such as \(x_{j_1}\), \(x_{j_1}^3\), \(x_{j_1}^2x_{j_2}\) \((j_1\neq j_2)\), and \(x_{j_1}x_{j_2}x_{j_3}\) \((j_1\neq j_2\neq j_3)\).
By solving the corresponding node-wise sparse regression problem, we recover both coupling forms of multiple orders and their associated interaction strengths.
Finally, the complete network topology is reconstructed by combining the identified self-dynamics \(\hat f\) and the learned coupling functions, following the topology-reconstruction step of the proposed framework.

\subsubsection{Impact of coupling strength}

Since the proposed framework is designed for weakly coupled networks, and since finite-difference differentiation inevitably introduces small numerical errors, it is important to examine how coupling strengths affect reconstruction accuracy.
Using the same Chen network setting as above, we systematically vary the pairwise coupling strength \(\sigma_1\) and the second-order coupling strength \(\sigma_2\).
For each parameter pair, we evaluate both the fitting accuracy of the hub-level effective coupling function and the topology reconstruction accuracy.

\begin{figure*}[htbp]
    \centering
    \begin{minipage}[t]{0.47\textwidth}
        \textbf{a}\par\vspace{1mm}
        \centering
        \includegraphics[width=\linewidth]{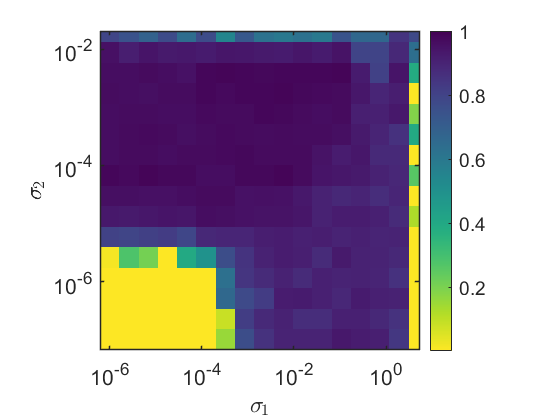}
    \end{minipage}
    \hspace{0.04\textwidth}
    \begin{minipage}[t]{0.47\textwidth}
        \textbf{b}\par\vspace{1mm}
        \centering
        \includegraphics[width=\linewidth]{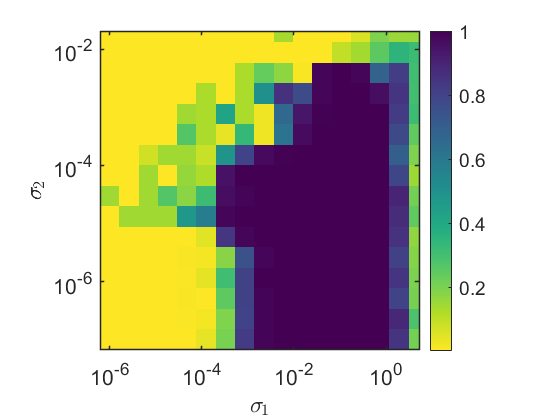}
    \end{minipage}

    \vspace{4mm}

    \begin{minipage}[t]{0.47\textwidth}
        \textbf{c}\par\vspace{1mm}
        \centering
        \includegraphics[width=\linewidth]{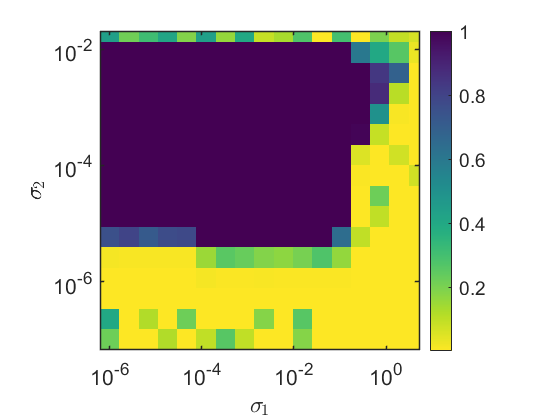}
    \end{minipage}
    \hspace{0.04\textwidth}
    \begin{minipage}[t]{0.47\textwidth}
        \textbf{d}\par\vspace{1mm}
        \centering
        \includegraphics[width=\linewidth]{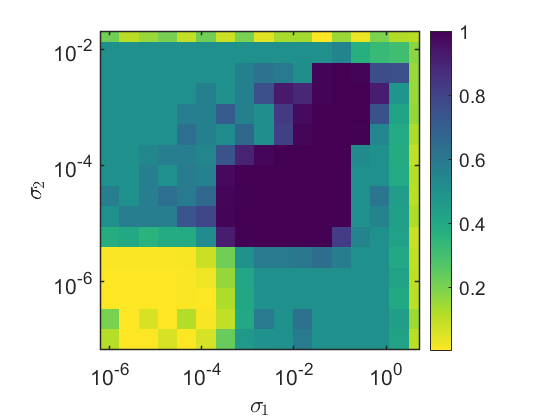}
    \end{minipage}

    \caption{
    Evaluation of how coupling strengths influence reconstruction performance.
    \textbf{a,} Fitting accuracy \(R^2\) of the hub-level effective coupling function.
    \textbf{b,} Reconstruction accuracy of pairwise interactions measured by the \(F_1\) score.
    \textbf{c,} Reconstruction accuracy of higher-order interactions measured by the \(F_1\) score.
    \textbf{d,} Overall \(F_1\) score combining different interaction orders.
    Accurate reconstruction is achieved within an effective region of the tested coupling-parameter space, which includes the representative weak-coupling settings considered in this study; parameter combinations outside this region reduce recoverability.
    }
    \label{fig2}
\end{figure*}

The overall reconstructability of the network depends on two factors.
The first factor is the fitting accuracy of the hub-level effective coupling function \(H(x)\), quantified by \(R^2\). As shown in Fig.~\ref{fig2}a, when both \(\sigma_1\) and \(\sigma_2\) are too weak, the amplitude of \(H(x)\) becomes comparable to the numerical noise introduced by derivative estimation.
Consequently, the coupling contribution is effectively buried, leading to low fitting accuracy.
Conversely, when either \(\sigma_1\) or \(\sigma_2\) is excessively large, one interaction component dominates the observed dynamics and obscures weaker components, thereby distorting the functional relation between \(H(x)\) and \(x\).
Within the effective region of the tested \((\sigma_1,\sigma_2)\) parameter grid, the effective coupling is well captured by a low-dimensional polynomial approximation.

The second factor is whether the reconstructed \(\hat f(x)\) and \(H(x)\) can be combined to recover the network topology.
As shown in Fig.~\ref{fig2}b, when \(\sigma_2\) is small, increasing \(\sigma_1\) improves pairwise reconstruction because the pairwise coupling becomes more distinguishable from numerical errors.
However, if \(\sigma_2\) increases while \(\sigma_1\) remains fixed, strong higher-order interactions interfere with pairwise identification and reduce the pairwise \(F_1\) score.
Similarly, Fig.~\ref{fig2}c shows accurate higher-order reconstruction within an effective part of the tested parameter grid, with performance deteriorating outside this region.
The overall performance in Fig.~\ref{fig2}d identifies an effective region within the tested coupling-parameter space. This region includes the representative weak-coupling settings considered here, in which the conditions required for the effective-coupling representation of higher-order interactions appear to hold in the simulations.

\subsubsection{Impact of network size}

To assess how network size affects reconstruction performance, we first use the second-order Chen network in Eq.~\eqref{eq:chen_second} and vary the number of nodes \(N\).
For each \(N\), higher-order networks are generated using the same procedure as above, while the Barabási--Albert attachment parameter is adjusted to ensure that node \(1\) acts as a hub.
At the pairwise level, node \(1\) connects to approximately \(20\)--\(30\%\), \(35\)--\(50\%\), or \(50\)--\(70\%\) of all nodes.
The number of second-order interactions involving node \(1\) is set to approximately \(1.5\) times its first-order degree.
These three settings are labeled as low, medium, and high hub-interaction density.
Throughout the second-order experiments, we fix \(\sigma_1=0.01\) and \(\sigma_2=0.001\).

We next extend the analysis to third-order networks.
The \(x\)-component is governed by
\begin{equation}
\label{eq:chen_third}
\begin{aligned}
\dot{x}_i &= \alpha (y_i - x_i) 
+ \sigma_{1} \sum_{j_1=1}^{N} a_{ij_1}^{(1)} (x_{j_1} - x_i)\\
&\quad+ \sigma_{2} \sum_{j_1=1}^{N} \sum_{j_2=1}^{N}
a_{ij_1j_2}^{(2)} (x_{j_1} x_{j_2} - x_i^{2}) \\
&\quad + \sigma_{3} \sum_{j_1=1}^{N} \sum_{j_2=1}^{N} \sum_{j_3=1}^{N}
a_{ij_1j_2j_3}^{(3)} (x_{j_1} x_{j_2} x_{j_3} - x_i^{3}),
\end{aligned}
\end{equation}
while the remaining variables follow the Chen dynamics in Eq.~\eqref{eq:chen_second}.
The coupling strengths are fixed at \(\sigma_1=0.01\), \(\sigma_2=0.001\), and \(\sigma_3=0.0001\).

\begin{figure*}[htbp]
    \centering
    \begin{minipage}[t]{0.47\textwidth}
        \textbf{a}\par\vspace{1mm}
        \centering
        \includegraphics[width=\linewidth]{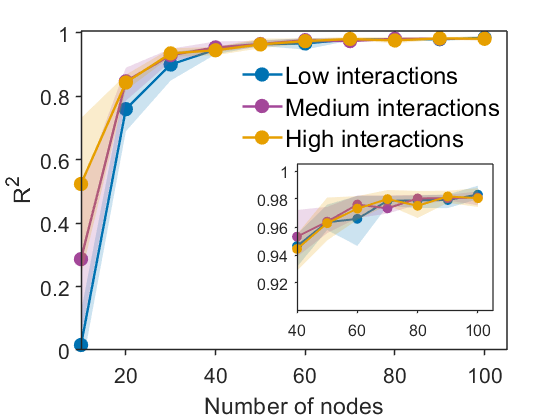}
    \end{minipage}
    \hspace{0.04\textwidth}
    \begin{minipage}[t]{0.47\textwidth}
        \textbf{b}\par\vspace{1mm}
        \centering
        \includegraphics[width=\linewidth]{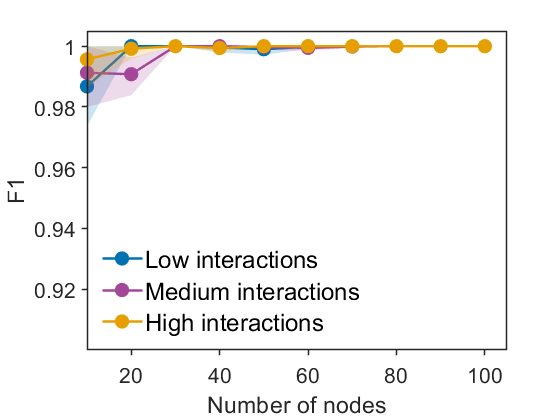}
    \end{minipage}

    \vspace{4mm}

    \begin{minipage}[t]{0.47\textwidth}
        \textbf{c}\par\vspace{1mm}
        \centering
        \includegraphics[width=\linewidth]{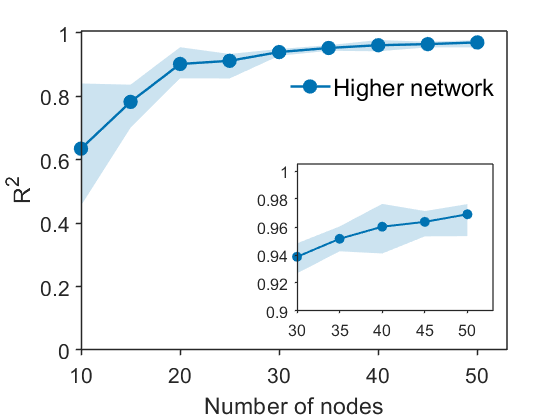}
    \end{minipage}
    \hspace{0.04\textwidth}
    \begin{minipage}[t]{0.47\textwidth}
        \textbf{d}\par\vspace{1mm}
        \centering
        \includegraphics[width=\linewidth]{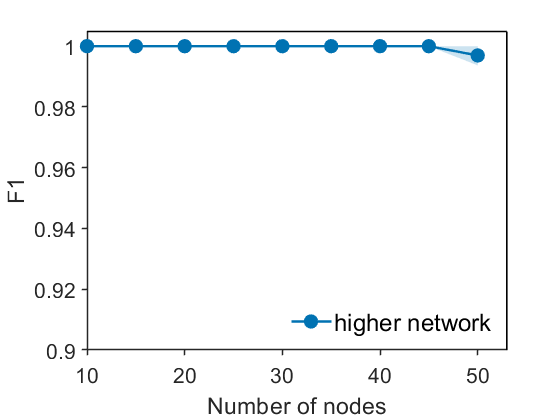}
    \end{minipage}

    \caption{
    Impact of network size on fitting and reconstruction performance for second- and third-order Chen networks.
    \textbf{a,} Hub-level fitting accuracy \(R^2\) for second-order Chen networks.
    \textbf{b,} Corresponding second-order topology reconstruction accuracy measured by the \(F_1\) score.
    \textbf{c,} Hub-level fitting accuracy \(R^2\) for third-order Chen networks.
    \textbf{d,} Corresponding third-order topology reconstruction accuracy measured by the \(F_1\) score.
    Stable reconstruction is maintained at larger network sizes and, where varied, at higher hub-neighborhood densities.
    }
    \label{fig_node_number}
\end{figure*}

As shown in Fig.~\ref{fig_node_number}a, the hub-level fitting accuracy for second-order networks increases with \(N\) and rapidly saturates near \(0.98\).
For a given \(N\), higher hub-neighborhood density accelerates this convergence, consistent with more accurate physical-measure averaging of the cumulative neighbor-driven input.
The corresponding reconstruction performance in Fig.~\ref{fig_node_number}b remains close to \(1\) across all tested network sizes and hub-density levels.

The third-order case shows a similar trend.
As shown in Fig.~\ref{fig_node_number}c, the hub-node fitting accuracy increases with \(N\) and approaches approximately \(0.96\) when \(N=50\).
Fig.~\ref{fig_node_number}d shows that the corresponding \(F_1\) scores remain above \(0.99\) across the tested network sizes.
At larger network sizes and higher hub-neighborhood densities, the proposed framework maintains stable effective-coupling fitting and accurate topological recovery. These results provide numerical support for the higher-order interaction reduction in both second- and third-order networks.

\subsection{Higher-order networks based on empirical topologies}

\begin{figure*}[htbp]
    \centering
    \begin{minipage}[t]{0.47\textwidth}
        \textbf{a}\par\vspace{1mm}
        \centering
        \includegraphics[width=\linewidth]{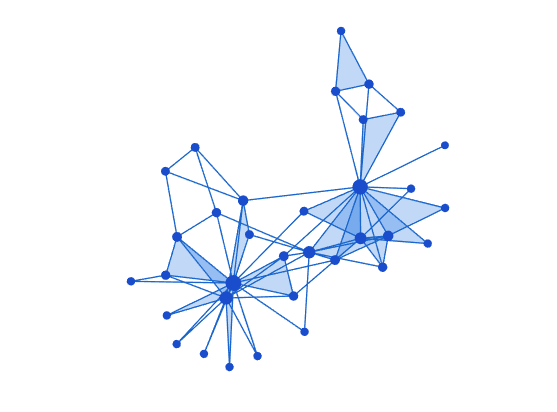}
    \end{minipage}
    \hspace{0.04\textwidth}
    \begin{minipage}[t]{0.47\textwidth}
        \textbf{b}\par\vspace{1mm}
        \centering
        \includegraphics[width=\linewidth]{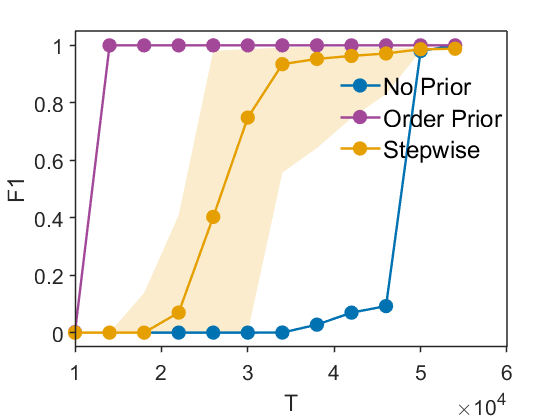}
    \end{minipage}

    \caption{
    Validation of the proposed framework on the Zachary Karate Club network.
    \textbf{a,} Higher-order interactions represented as directed 2-simplices on the empirical pairwise topology, with simplex orientations omitted for clarity.
    \textbf{b,} Reconstruction accuracy as a function of time-series length under no-prior, stepwise-filtering, and order-prior strategies.
    Reducing the candidate library through structural priors or stepwise filtering significantly improves reconstruction efficiency.
    }
    \label{fig5}
\end{figure*}

To further assess the practical applicability of the proposed framework, we conduct simulations on the Zachary Karate Club network, a standard social-network benchmark consisting of \(34\) nodes and \(78\) pairwise edges.
From this base topology, we generate \(120\) directed 2-simplices to model higher-order interactions, as shown in Fig.~\ref{fig5}a.
The node dynamics follow the Chen model in Eq.~\eqref{eq:chen_second}, maintaining consistency with the synthetic-network analyses.

This experiment evaluates the sensitivity of the method to time-series length.
The required sample size depends strongly on the dimension of the candidate function library.
Therefore, reducing the library size can substantially lower data requirements.
We compare three reconstruction strategies.
In the no-prior strategy, all considered candidate terms are retained.
In the order-prior strategy, the maximum interaction order \(D\) is assumed to be known, so irrelevant higher-order terms are excluded.
For example, in a second-order network, terms such as \(x_{j_1}\) and \(x_{j_1}^2x_{j_2}\) are retained, whereas \(x_{j_1}^3\) and \(x_{j_1}x_{j_2}x_{j_3}\) are discarded.
In the stepwise strategy, lower-order interactions are first inferred from short time windows and then used to restrict the candidate set for higher-order reconstruction.

All data before \(t=100\) are excluded to remove transient behavior and provide an additional settling buffer; reconstruction begins at \(t=100\).
The integration step size is fixed at \(0.01\).
As shown in Fig.~\ref{fig5}b, without prior knowledge of the network order, the \(F_1\) score requires a long observation window to approach \(1\).
In contrast, when the interaction order is known a priori, the reduced candidate space allows the \(F_1\) score to reach \(1\) with much shorter time series.
The stepwise strategy achieves intermediate performance, requiring substantially fewer samples than the no-prior baseline.
These results show that the framework can be applied to empirical network topologies and that weak structural priors can improve reconstruction efficiency without changing the core inference procedure.

\subsection{Cat cortical topology with heterogeneous node dynamics}

\begin{figure*}[htbp]
    \centering
    \begin{minipage}[t]{0.47\textwidth}
        \textbf{a}\par\vspace{1mm}
        \centering
        \includegraphics[width=\linewidth]{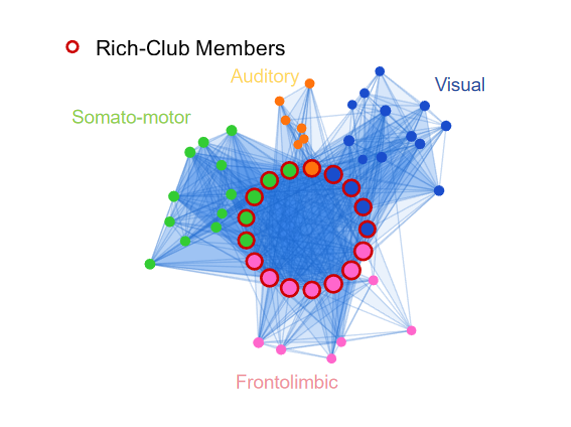}
    \end{minipage}
    \hspace{0.04\textwidth}
    \begin{minipage}[t]{0.47\textwidth}
        \textbf{b}\par\vspace{1mm}
        \centering
        \includegraphics[width=\linewidth]{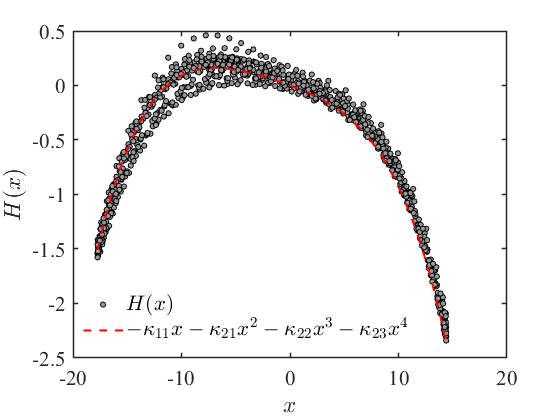}
    \end{minipage}

    \caption{
    Numerical validation on the cat cortical topology with heterogeneous node dynamics.
    \textbf{a,} Rich-club nodes connect the visual, auditory, somatosensory--motor, and frontolimbic subsystems, participating in intra- and inter-system higher-order couplings.
    \textbf{b,} Reconstructed effective coupling function \(H(x)\) for a representative hub node and its polynomial fit.
    The fitted coefficients combine coupling strengths and participation counts of distinct interaction types, enabling inference of multi-system higher-order coupling patterns.
    }
    \label{fig6}
\end{figure*}

To assess the framework on a heterogeneous network based on an empirical topology, we analyze the cat cortical network, whose mesoscale connectivity has been mapped using anatomical tracing data~\cite{scannell1993connection,scannell1995analysis}.
The network comprises \(53\) cortical regions grouped into four major functional modules: visual, auditory, somatosensory--motor, and frontolimbic regions.
As shown in Fig.~\ref{fig6}a, we adopt the rich-club organization from Ref.~\cite{eroglu2020revealing}, where rich-club nodes form a densely interconnected backbone linking the four functional modules.
On top of this structure, we introduce \(156\) directed 2-simplices to represent higher-order interactions.

Given the functional heterogeneity and non-scale-free organization of the cat cortex, we perform reconstruction under node-specific dynamical constraints.
The visual and auditory regions follow the Lorenz system,
\[
\dot{x}=10(y-x),\quad 
\dot{y}=28x-y-xz,\quad 
\dot{z}=xy-\tfrac{8}{3}z.
\]
The somatosensory--motor region follows the Rössler system,
\[
\dot{x}=-y-z,\quad 
\dot{y}=x+0.2y,\quad 
\dot{z}=0.2+z(x-5.7).
\]
The frontolimbic region follows the Chen system,
\[
\dot{x}=35(y-x),\quad 
\dot{y}=-7x-xz+28y,\quad 
\dot{z}=xy-3z.
\]

Pairwise interactions are modeled as diffusion-like couplings \(x_{j_1}-x_i\), while nonlinear cooperative effects between subsystems are represented by second-order couplings:
\begin{equation}
\begin{aligned}
H_i^{(2)}(x)
&= \sigma_2 \sum_{j_1,j_2} a_{ij_1j_2}^{(2)}
  0.1^{(2-\delta_{ij_1}-\delta_{ij_2})} \\
&\quad \times
  \left(
  x_{j_1}^{(2-\delta_{ij_1})}
  x_{j_2}^{(2-\delta_{ij_2})}
  - x_i^{(4-\delta_{ij_1}-\delta_{ij_2})}
  \right),
\label{eq:high_order_adapt}
\end{aligned}
\end{equation}
where \(\delta_{ij_r}=1\) if nodes \(i\) and \(j_r\) belong to the same subsystem, and \(\delta_{ij_r}=0\) otherwise, with \(j_r\neq i\).
This definition assigns stronger effective weights to intra-system couplings than to inter-system couplings, reflecting the modular organization of the cortical network.

Under weak coupling, each subsystem maintains the statistical stability of its intrinsic dynamics.
Applying the effective-coupling representation to this multi-system setting, the collective influence on a hub node can be approximated as
\begin{equation}
\begin{aligned}
H(x_i(t)) &=
\sigma_1 a_{ii}^{(1)}h^{(1)}(x_i(t))
+ \sum_{m} \sigma_{2m} a_{iii}^{(2)} h^{(2)}_m(x_i(t),x_i(t)) \\
&\quad + C + \varepsilon_i(t),
\label{eq:multi_system_reduced_final}
\end{aligned}
\end{equation}
where \(m\) denotes distinct types of second-order interactions, \(C\) represents the aggregated effect of cross-system coupling, and \(\varepsilon_i(t)\) is a bounded fluctuation induced by finite sampling of neighbor states.
The repeated sampling across chaotic subsystems is consistent with an approximately time-independent contribution \(C\), supporting the use of the reduced representation in this heterogeneous setting.

Figure~\ref{fig6}b shows the reconstructed effective coupling function for a representative hub node.
The fitted curve takes the form
\begin{equation}
H(x) = -\kappa_{11}x - \kappa_{21}x^2 - \kappa_{22}x^3 - \kappa_{23}x^4,
\label{eq:cat_effective_coupling}
\end{equation}
where \(\kappa_{11}=\sigma_1 s_1\) and \(\kappa_{2m}=\sigma_{2m}s_{2m}\) for \(m=1,2,3\).
Here, \(s_1\) and \(s_{2m}\) represent the participation counts of first-order and different types of second-order couplings, respectively.
By using the mapped coupling curve \(H(x)\) to construct the reduced candidate library and solving the resulting regression problem, the framework recovers effective coupling coefficients and active pairwise and higher-order interaction patterns in the heterogeneous cortical network.

\section{Conclusion}

This study identifies a higher-order interaction reduction in chaotic networks: under weak coupling and dense local connectivity, the many joint multi-node inputs received by a hub can be represented collectively by a compact effective coupling governed primarily by the hub state. Building on this reduction, we develop a reconstruction framework that starts solely from continuous-valued time series, identifies intrinsic node dynamics, and recovers the order, direction and functional form of multi-node interactions.

In randomly generated higher-order Chen networks, the framework stably fits hub-level effective coupling functions and accurately reconstructs interactions of different orders within an effective region of the tested coupling-parameter space, which includes the representative weak-coupling settings considered in this study. When the coupling is too weak, the interaction signal is susceptible to finite-sample and numerical differentiation errors; when it is too strong, the conditions underlying the reduction gradually cease to hold and different interaction orders become harder to separate. Reconstruction remains stable at the larger network sizes and higher hub-neighborhood densities examined. Experiments on higher-order networks constructed from the pairwise topologies of the Zachary Karate Club network and the cat cortical network further demonstrate the applicability of the framework to observed topologies, heterogeneous node dynamics and rich-club higher-order couplings.

The mean-field step in the derivation relies on physical-measure averaging of neighbor-driven sums. Although this treatment has been rigorously justified for certain classes of chaotic systems~\cite{pereira2020heterogeneously}, a general proof for arbitrary chaotic systems remains unavailable. Nevertheless, numerical simulations across different node dynamics and network structures demonstrate that the reduction remains applicable within the parameter regimes examined in this study.

Despite its advantages, the current framework has several clear limitations.
First, the method requires nodes to exhibit sufficient distinguishability in their intrinsic dynamics or coupling environments, so that self-dynamics and coupling-induced effects can be separated from the observed time series.
When hub nodes and their corresponding node dynamics are partly known a priori, this requirement can be relaxed, but in fully unknown systems with weakly distinguishable nodes the reconstruction may become less reliable.
Second, although the reduced candidate library alleviates the curse of dimensionality, the number of possible higher-order terms still grows rapidly with both the interaction order and the network size.
This imposes practical limits on the highest interaction order that can be inferred in very large systems.
Third, the current formulation assumes that sufficiently informative time series are available and that the relevant state variables are observable.
In systems with strong observational noise, hidden variables or partial measurements, additional regularization or data-assimilation strategies may be required.

Several extensions are possible.
Prior structural information, when available, can be incorporated to further reduce the candidate space, as illustrated by the order-prior and stepwise strategies.
The framework can also be combined with more advanced sparse regression, weak-form identification or ensemble-based approaches to improve robustness to noise and numerical differentiation errors.
For very large networks, preliminary screening of candidate neighbors, motif-based filtering or parallel node-wise reconstruction may further reduce computational cost.
Overall, the proposed framework provides a scalable and physically interpretable route to dynamical identification and multi-order topology recovery in higher-order chaotic networks.

\section*{Funding}
This work was supported by the National Natural Science Foundation of China under Grants 62573301 and 62503335, the Shenzhen University Startup Grants, and Guangdong Provincial Key Laboratory of Visual Media and Multidimensional Intelligence (2026B1212010012).

\section*{CRediT authorship contribution statement}
Jiahao He: Methodology, Software, Validation, Formal analysis, Investigation, Visualization, Writing -- original draft. Xiaoqun Wu: Conceptualization, Methodology, Supervision, Project administration, Funding acquisition, Writing -- review and editing. Ziye Fan: Methodology, Formal analysis, Writing -- review and editing. Jianwen Feng: Supervision, Resources, Writing -- review and editing. Tingwen Huang: Conceptualization, Supervision, Writing -- review and editing.

\section*{Declaration of competing interest}
The authors declare that they have no known competing financial interests or personal relationships that could have appeared to influence the work reported in this paper.

\section*{Data availability}
The data and code supporting the findings of this study are available from the corresponding author upon reasonable request.

\bibliographystyle{elsarticle-num}
\bibliography{wenxian}

\end{document}